\documentclass{aa}
\usepackage{natbib}
\bibpunct{(}{)}{;}{a}{}{,}
\usepackage{graphicx}
\usepackage{txfonts}
\usepackage{longtable,lscape}
\def\lta{\ifmmode{\,\mathrel{\mathpalette\@versim<\,}}
   \else{$\,\mathrel{\mathpalette\@versim<}\,$}\fi} 
\begin{document}
   \title{A new Planetary Nebula in the outer reaches of the Galaxy}

   \author{K. Viironen\inst{1,2,3}
	  \and
          A. Mampaso\inst{1,4}
          \and          
          R. L. M. Corradi\inst{1,4}
	  \and
          J. E. Drew\inst{5}
          \and
          D. J. Frew\inst{6}
          \and
          C. Giammanco\inst{1,4}
          \and
          R.~Greimel\inst{7}
          \and
	  T. Liimets\inst{8,9}
          \and
          J. E. Lindberg\inst{8,10}
          \and
          M. Rodr{\'{i}}guez\inst{11}
          \and
	  L. Sabin\inst{12}
          \and
	  S.~E.~Sale\inst{13,5,14,15}
          \and
          P.~A.~Wilson\inst{8,16}
          \and
          A. Zijlstra\inst{17}
          }

   \offprints{K. Viironen}

   \institute{Instituto de Astrof{\'{i}}sica de Canarias (IAC), C/V{\'{i}}a L\' actea s/n,
              38200 La Laguna, Tenerife, Spain\\
         \and
         Centro Astron\'omico Hispano Alem\'an, Calar Alto, C/Jes\'us Durb\'an Rem\'on 2-2, E-04004 Almeria, Spain \\
          \and
          Centro de Estudios de F\'isica del Cosmos de Arag\'on (CEFCA), C/General Pizarro 1-1, E-44001 Teruel, Spain \\
          \email{kerttu@cefca.es}\\
          \and
          Departamento de Astrof\'{i}sica, Universidad de La Laguna, Av. Astrof\'isico Francisco S\'anchez, s/n, 38206 La Laguna, Spain\\
          \and
   	 Centre for Astrophysics Research, University of Hertfordshire, College Lane, Hatfield AL10 9AB, UK\\       
         \and
         Department of Physics and Astronomy, Macquarie University, North Ryde, NSW 2109, Australia\\
         \and 	 
	 Institut f\"ur Physik, Karl-Franzens Universit\"at Graz, Universit\"atsplatz 5,8010 Graz, Austria\\
         \and
         Nordic Optical Telescope, Ap. 474, E-38700, S/C de La Palma, S/C de Tenerife, Spain\\
         \and
         Tartu Observatory, T\~oravere, 61602, Estonia \\
         \and
         Centre for Star and Planet Formation, Natural History Museum of
Denmark, University of Copenhagen, \O ster Voldgade 5-7, DK-1350 Copenhagen, Denmark\\
	 \and
          Instituto Nacional de Astrof{\'{i}}sica, \'Optica y Electr\'onica (INAOE), Apdo Postal 51 y 216, 72000 Puebla, Mexico\\
          \and
	 Instituto de Astronom\'{i}a UNAM, Km. 103 Carretera Tijuana-Ensenada, C.P. 22860, Ensenada, Baja California, Mexico\\
	 \and
	 Imperial College London, Blackett Laboratory, Prince Consort Road, London SW7 2AZ, UK\\
         \and
         Departamento de F\'{i}sica y Astronom\'{i}a, Facultad de Ciencias, Universidad de Valpara\'{i}so, Ave. Gran Breta\~na 1111, Playa Ancha, Casilla 53,
Valpara\'{i}so, Chile\\
         \and
         Departamento de Astronom\'{i}a y Astrof\'{i}sica, Pontificia Universidad Cat\'olica de Chile, Av. Vicu\~na Mackenna 4860, Casilla 306, Santiago 22,
Chile\\
         \and
Institute of Theoretical Astrophysics, University  of Oslo, P.O. Box 1029 Blindern, N-0315 Oslo, Norway\\
         \and
	 Jodrell Bank Centre for Astrophysics, Alan Turing Building, University of Manchester, Manchester, M13 9PL, UK\\
}

   \date{}


  \abstract
 {} 
 {A proper determination of the abundance gradient in the Milky Way 
requires the observation of objects at large galactiocentric distances. 
With this aim, we are exploring 
the planetary nebula population towards the Galactic Anticentre. In this article, 
the discovery and physico-chemical study of a new planetary nebula towards the 
Anticentre direction, IPHASX J052531.19+281945.1 (PNG 178.1-04.0), is presented.}
    {The planetary nebula was discovered from the IPHAS survey. Long-slit follow-up spectroscopy was carried out to confirm its planetary nebula nature and to calculate its physical and chemical characteristics.}
    {The newly discovered planetary nebula turned out to be located at a very large galactocentric distance (D$_{GC}=20.8\pm3.8$ kpc), larger than any previously known planetary nebula with measured abundances. 
Its relatively high oxygen abundance (12+log(O/H) = 8.36$\pm$0.03) supports a flattening of the Galactic abundance gradient at large galactocentric distances rather than a linearly decreasing gradient.}
    {}


   \maketitle
%

\section{Introduction}

The INT Photometric H-Alpha Survey of the Northern Galactic Plane
\cite[IPHAS,][]{drew05,gonzalez-solares08} is significantly increasing
the number of detected Galactic planetary nebulae (PNe). \citet{viironen09b}
published a list of 781 new IPHAS PN candidates, whereas the discovery
and detailed study of five new IPHAS PNe were presented by
\citet{viironen09a}.

PNe towards the Galactic Anticentre are of special interest as this
is where the objects with the largest galactocentric distances are most easily found and studied. Finding new distant PNe will make a crucial
contribution to the knowledge of the Galactic abundance gradient, since
their rich emission line spectra can yield accurate abundances. The
abundance gradient is a topic that is still under debate \citep[see
  e.g.][and references therein]{maciel09}. In addition to the shortage
of known objects at large galactocentric distances, the main problem
is the difficulty of reliably measuring PN distances \citep[see][for
  example]{kwok00}. The new faint (and therefore potentially distant)
IPHAS Anticentre PN candidates combined with new methods for distance
determination \citep{sale09,frew06} can make a significant
contribution to the problem.

\begin{figure*}
   \centering
   \includegraphics[width=0.3\linewidth,angle=-90]{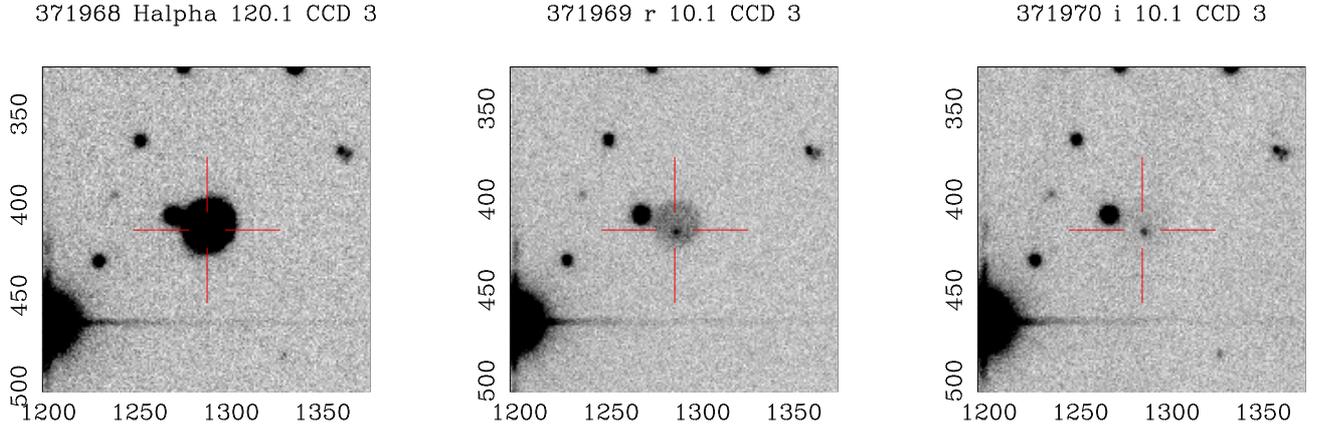}
      \caption{IPHAS H$\alpha$, $r^{\prime}$ and $i^{\prime}$ discovery images
        of IACPN, from left to right. The
        image sizes are $1'\times1'$, North is up and East to the
        left.}
\label{fig:1}
\end{figure*}

The location of the newly discovered PN IPHASX J052531.19+281945.1, at
only four degrees from the direction of the Galactic Anticentre,
motivated us to study this object further. In Sect.~\ref{obs} the
photometric and spectroscopic observations are presented; in
Sect.~\ref{analysis} a physical and chemical analysis is carried out; in Sect.~\ref{dist} its distance
is studied; in Sect.~\ref{disc} a discussion is provided, and in Sect.~\ref{conc} the main conclusions are presented.

\section{Observations and data reduction}\label{obs}

The PN was discovered by visually inspecting the
IPHAS H$\alpha$ $-$ $r^{\prime}$ mosaics of the Anticentre region. The semi-automated searching
method described in \citet{viironen09b} missed the object due to its
relatively large size ($\sim$ 10\arcsec). The IAU approved name for this IPHAS object is, after its coordinates, IPHASX J052531.19+281945.1 (X for extended) whereas the IAU common PN name would be PNG 178.1-04.0. To abbreviate, in the following we will use the acronym IACPN (IPHAS Anti-Centre Planetary Nebula).

\subsection{IPHAS imaging}

The IPHAS observations of IACPN were carried out on November 2003 using
the 2.5m Isaac Newton Telescope and its Wide Field Camera under seeing condition of 1.3$\arcsec$ FWHM
in the H$\alpha$ images. Fig.~\ref{fig:1} presents the H$\alpha$, $r^{\prime}$ and $i^{\prime}$ images of the object. The
exposure times are 120 s for H$\alpha$ and 10 s for $r^{\prime}$ and
$i^{\prime}$ images.
More details about IPHAS observations and the data reduction are given
in \cite{drew05} and \cite{gonzalez-solares08}.

IACPN shows a roughly spherical morphology. It is
hardly visible in the IPHAS $i^{\prime}$ filter
(see Fig.~\ref{fig:1}). No objects are listed in SIMBAD within almost
5\arcmin~from the position of IACPN.

\subsection{SPM 2.1-m spectroscopy}\label{sec:spm}

The first spectroscopic observations of IACPN were carried out
at the Observatorio de San Pedro M\'artir (Mexico), and were presented
by \cite{mampaso05} in a conference paper. Four 20 min exposures in
the red and five 20 min exposures in the blue were obtained through a narrow (1$\arcsec$) slit using the 2.1-m telescope and the B\&Ch spectrograph
(2 \AA~pix$^{-1}$ dispersion, 3 \AA~resolution, from 3720 to 7470 \AA). In addition, two 10 min exposures in
the blue were obtained with a very wide slit in order to include the
whole nebula and measure its total H$\beta$ flux. Furthermore, the
MEZCAL echelle spectrograph was used with the same
telescope to obtain a high dispersion spectrum (0.1 \AA~pix$^{-1}$) around
the H$\alpha$ line.

The MEZCAL spectrum allowed to measure the expansion and radial
velocities of the nebula: the $\sim$10\arcsec diameter shell has a radial
velocity of $v_{LSR} = 13.5$ km s$^{-1}$ and expands at 17 km s$^{-1}$ in the [N\,{\sc ii}] 6583 \AA~line \citep{mampaso05}. From the B\&Ch
spectra the PN nature of the nebula was confirmed and its absolute
H$\beta$ flux was measured, giving F(H$\beta)= 3.34 \times
10^{-14}$~erg~cm$^{-2}$~s$^{-1}$. Preliminary chemical abundances were
presented by \cite{mampaso05}.

\subsection{WHT + ISIS spectroscopy}\label{sec:wht}

The William Herschel Telescope (WHT) + ISIS spectrograph observations
of IACPN were carried out in service mode on April 6, 2007. The seeing
was $\sim 0.9\arcsec$ and the weather conditions were good, except for
the presence of light cirrus during the second half of the
night. Three exposures of 20 min of the target were obtained in both
the blue (3570-5115 \AA; dispersion 0.8 \AA~pix$^{-1}$, resolution 3.2
\AA) and the red (5540-10500 \AA, 1.8 \AA~pix$^{-1}$ dispersion, 6.5
\AA~resolution) arms with the slit located at P.A. = 79$^o$, i.e. at
the parallactic angle corresponding to the moment of observation. The
slit width was 1$\arcsec$. In addition, bias, lamp flats, and arc
exposures were obtained. Unfortunately, no suitable observations of
spectrophotometric standards are available for that night, as all
  standard stars were observed with a different instrumental
  configuration than our target. For this reason, a spectrum of the
  standard star BD+332642 \citep{oke90}, obtained on April 2, 2007
  with the same setup as for the nebula, was used to flux calibrate
  its spectrum.  Nevertheless, the blue and red observations of this standard star
  were obtained
  at different times of the night, and therefore under different
  weather conditions. This prevents a precise matching of the blue and
  red sides of the WHT spectrum, and therefore, an accurate 
  determination of $A_V$ using the H$\alpha$/H$\beta$ flux ratio. For this
  reason, additional spectra of the nebula and standard star were acquired as described below. 
Finally, and given that there is calibration data available for BD+33 2642 only up
to 9200 \AA~\citep{oke90}, the fluxes for lines at longer wavelengths
will not be considered. The data reduction and calibration were
carried out using the standard IRAF routines for longslit
spectroscopy.

\begin{table}
\caption{\label{tab:wht} WHT+ISIS observed line fluxes F (Obs.) normalised to  F(H$\beta$) = 100 in the blue section of the spectrum (separated by a horizontal line) and to  F(H$\alpha$) = 100 in the red section. See text for details about the normalisation of the dereddened fluxes, F (deredd). The wavelengths are in units of \AA~and the total errors in the flux measurements and the propagated errors in the dereddened flux are given within brackets in percentage.}
\centering
 \begin{tabular}{@{}lllll@{}}
 \hline \hline
Line, $\lambda$ & $\lambda$~Obs. & F (Obs.) & F (deredd)\\
 \hline
{}[O\,{\sc ii}] 3726.03+3728.82 & 3727.8             &213.7 (2) & 379.9 (7) \\  
H\,{\sc i} 3797.90 &3797.4                           &2.8 (15) &  4.8 (16) \\ 
H\,{\sc i} 3835.39 &3835.3                           &4.4 (10) &  7.5 (11) \\ 
{}[Ne\,{\sc iii}] 3868.75 &3868.9                    &57.1 (2) & 94.8 (6) \\ 
He\,{\sc i} 3888.65 + H\,{\sc i} 3889.05 & 3889.0    &13.3 (4) & 21.9 (7) \\ 
{}[Ne\,{\sc iii}] 3967.46+H\,{\sc i}3970.07 & 3968.4 & 28.1 (3) & 44.4 (6) \\ 
H$\delta$ 4101.74 &4101.8                            &17.0 (4) & 25.2 (6) \\ 
H$\gamma$ 4340.47 &4340.7                            &36.8 (2) & 47.9 (4) \\ 
{}[O\,{\sc iii}] 4363.21 &4363.2                     &5.4 (8) &  6.9 (8) \\ 
He\,{\sc i} 4471.50 &4471.7                          &4.1 (11) &  5.0 (12) \\ 
He\,{\sc ii} 4685.68 &4685.7                         &9.6 (7) & 10.5 (7) \\ 
H$\beta$ 4861.33 &4861.6                             &100.0 (2) & 100.0 (2) \\ 
{}[O\,{\sc iii}] 4958.91 &4959.1                     &238.6 (2) & 227.7 (2) \\ 
{}[O\,{\sc iii}] 5006.84 &5007.1                     &728.9 (1) & 680.0 (2) \\  
\hline
{}[N\,{\sc ii}] 5754.64 &5754.1                      & 0.8 (11) &  3.1 (12) \\
He\,{\sc i} 5875.64 &5875.1                          & 4.4 (3) & 15.9 (5) \\
{}[O\,{\sc i}] 6300.30 &6299.9                       & 8.1 (2) & 24.9 (7) \\
{}[S\,{\sc iii}] 6312.10 &6312.0                     & 0.4 (16) &  1.2 (18) \\ 
{}[O\,{\sc i}] 6363.78 &6363.4                       & 2.7 (3) &  8.1 (7) \\
{}[N\,{\sc ii}] 6548.03 &6547.5                     & 18.6 (1) & 53.4 (7) \\ 
H$\alpha$ 6562.82 &6562.2                           & 100.0 (1) & 286.0 (7) \\ 
{}[N\,{\sc ii}] 6583.41 &6582.8                     & 55.5 (1) & 157.7 (7) \\ 
He\,{\sc i} 6678.15 &6677.6                         & 1.7 (5) &  4.7 (9) \\
{}[S\,{\sc ii}] 6716.47 &6715.9                      & 7.6 (1) & 20.9 (8) \\
{}[S\,{\sc ii}] 6730.85 &6730.4                      & 6.2 (2) & 16.8 (8) \\
He\,{\sc i} 7065.28 &7064.8                          & 1.5 (3) &  3.8 (9) \\
{}[Ar\,{\sc iii}] 7135.78 &7135.4                    & 3.6 (2) &  8.9 (9) \\
C\,{\sc ii} 7236.42 &7235.0                          & 0.2 (28) &  0.5 (30) \\
He\,{\sc i} 7281.35 &7280.8                          & 0.3 (23) &  0.8 (24) \\
{}[O\,{\sc ii}] 7318.39+7319.99 &7319.5              & 2.4 (3) &  5.7 (10) \\
{}[O\,{\sc ii}] 7329.66+7330.73 &7329.9              & 2.1 (4) &  5.0 (10) \\
{}[Ar\,{\sc iii}] 7751.10 &7750.7                    & 1.0 (9) &  2.1 (14) \\
{}[C\,{\sc i}] 8727.13 &8727.1                       & 1.0 (8) &  1.8 (15) \\
H\,{\sc i} 8750.47 &8750.7                           & 0.5 (14) &  1.0 (18) \\
H\,{\sc i} 8862.79 &8861.9                           & 0.6 (21) &  1.0 (24) \\
H\,{\sc i} 9014.91 &9014.8                           & 1.0 (18) &  1.8 (22) \\
{}[S\,{\sc iii}] 9068.9 &9068.7                      & 4.5 (2) &  7.8 (13) \\
\hline                                               
\end{tabular}                                       
\end{table}                                         
                                                    
\begin{table}                                       
\caption{\label{tab:not} As in Table 1 for the NOT spectrum normalised to 
F(H$\beta$)=100.}
\centering
 \begin{tabular}{@{}llll@{}}
 \hline \hline
Line, $\lambda$ & $\lambda$~Obs. & F (Obs.)& F (deredd)  \\
 \hline
{}[Ne\,{\sc iii}] 3868.75 & 3868.7 & 61.8 (9) & 101.8 (11) \\ 
He\,{\sc i} 3888.65 + H\,{\sc i} 3889.05 & 3889.2 & 13.8 (25) & 22.4 (26) \\ 
{}[Ne\,{\sc iii}] 3967.46+H\,{\sc i}3970.07 & 3967.5 & 29.6 (10) & 46.4 (12) \\ 
H$\delta$ 4101.74 & 4101.5 & 19.3 (10) & 28.4 (11) \\ 
H$\gamma$ 4340.47 & 4340.2 & 37.0 (5) & 48.0 (6) \\ 
He\,{\sc ii} 4685.68 & 4685.2 & 11.3 (11) & 12.3 (11) \\ 
H$\beta$ 4861.33 & 4860.9 & 100.0 (2) & 100.0 (2) \\ 
{}[O\,{\sc iii}] 4958.91 & 4958.4 & 243.8 (2) & 232.9 (2) \\ 
{}[O\,{\sc iii}] 5006.84 & 5006.4 & 742.1 (1) & 693.2 (2) \\ 
{}[N\,{\sc ii}] 5754.64 & 5754.8 &  4.4 (21) &  3.0 (22) \\ 
He\,{\sc i} 5875.64 & 5874.8 & 22.1 (5) & 14.5 (7) \\ 
{}[O\,{\sc i}] 6300.30 & 6299.8 & 41.7 (5) & 23.7 (9) \\ 
{}[S\,{\sc iii}] 6312.10 & 6310.6 &  2.5 (57) &  1.4 (58) \\ 
{}[O\,{\sc i}] 6363.78 & 6362.7 & 14.5 (7) &  8.1 (11) \\ 
{}[N\,{\sc ii}] 6548.03 & 6546.9 & 96.5 (4) & 51.0 (9) \\ 
H$\alpha$ 6562.82 & 6561.8 & 553.9 (4) & 291.3 (9) \\ 
{}[N\,{\sc ii}] 6583.41 & 6582.5 & 311.4 (4) & 162.7 (9) \\ 
He\,{\sc i} 6678.15 & 6677.3 &  8.4 (18) &  4.3 (20) \\ 
{}[S\,{\sc ii}] 6716.47 & 6715.2 & 37.7 (5) & 19.0 (10) \\ 
{}[S\,{\sc ii}] 6730.85 & 6729.5 & 29.4 (5) & 14.7 (11) \\
\hline
\end{tabular}
\end{table}

\subsection{NOT + ALFOSC spectroscopy}

A good measurement of the reddening of the nebula, $A_V$, is
  important for its distance determination (see Sec. \ref{dist}) and additional spectra of IACPN were 
  secured in service mode on September 6, 2010 using the ALFOSC
  instrument at the Nordic Optical Telescope.  These spectra were 
  obtained so as to cover the wavelength range from $\sim$3800 \AA\ to
  6800 \AA, i.e. to include simultaneously the main Balmer lines
  (H$\delta$, H$\gamma$, H$\beta$, and H$\alpha$). The night was
photometric and the seeing was $0.7\arcsec$. Two 20 min exposures were
obtained. The spectral dispersion was 1.5 \AA~pix$^{-1}$ and
resolution 6 \AA. The slit was positioned according to the parallactic
angle at the moment of observation (P.A. = 102$^o$) and the slit width
used was 1$\arcsec$. Bias, lamp flats, and arc exposures were
obtained, and the standard star G191-B2B was observed. The data
reduction and calibration were carried out using the standard IRAF
routines for longslit spectroscopy.

\begin{figure}
   \centering
   \includegraphics[width=\linewidth,angle=0]{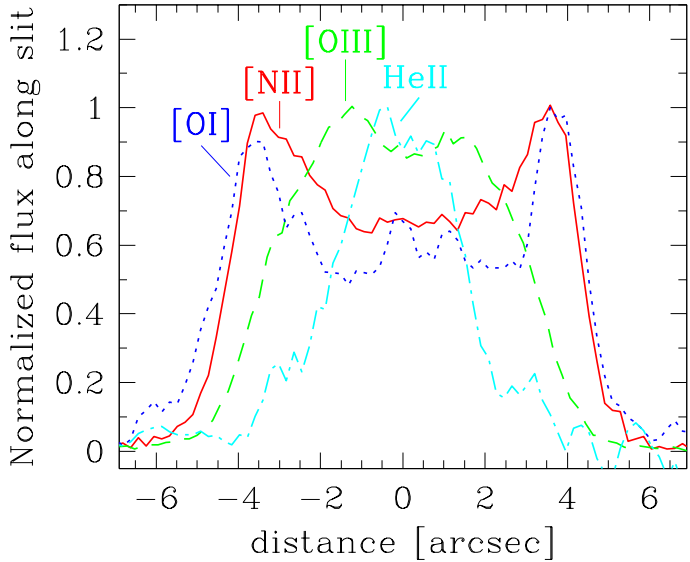}
      \caption{Spatial profiles extracted from the NOT spectrum for representative lines from ions with different ionisation potentials.}
\label{fig:2}
\end{figure}

\section{Spectroscopic analysis}\label{analysis}

The spatial distribution of the emission from several lines in the NOT
spectrum is presented in Fig.~\ref{fig:2}, and shows that IACPN
exhibits a strong ionisation stratification, with higher excitation
emissions coming from regions progressively closer to the centre. For
this reason, care has to be exercised when comparing line fluxes from
the WHT and NOT spectra. The difference in slit P.A. for both datasets
is relatively small (23$^o$) and the spatial profiles for the WHT
spectrum (not shown here) are very similar to those of the NOT
(Fig.~\ref{fig:2}) a likely consequence of the round symmetry of the
object. On the other hand, in the WHT data the slit passes through the
star partly overlapping the nebula on its eastern side (see
Fig.~\ref{fig:1}), and special care was taken to extract this star
from the nebular spectrum.

\begin{figure*}
   \centering
   \includegraphics[width=\linewidth,angle=0]{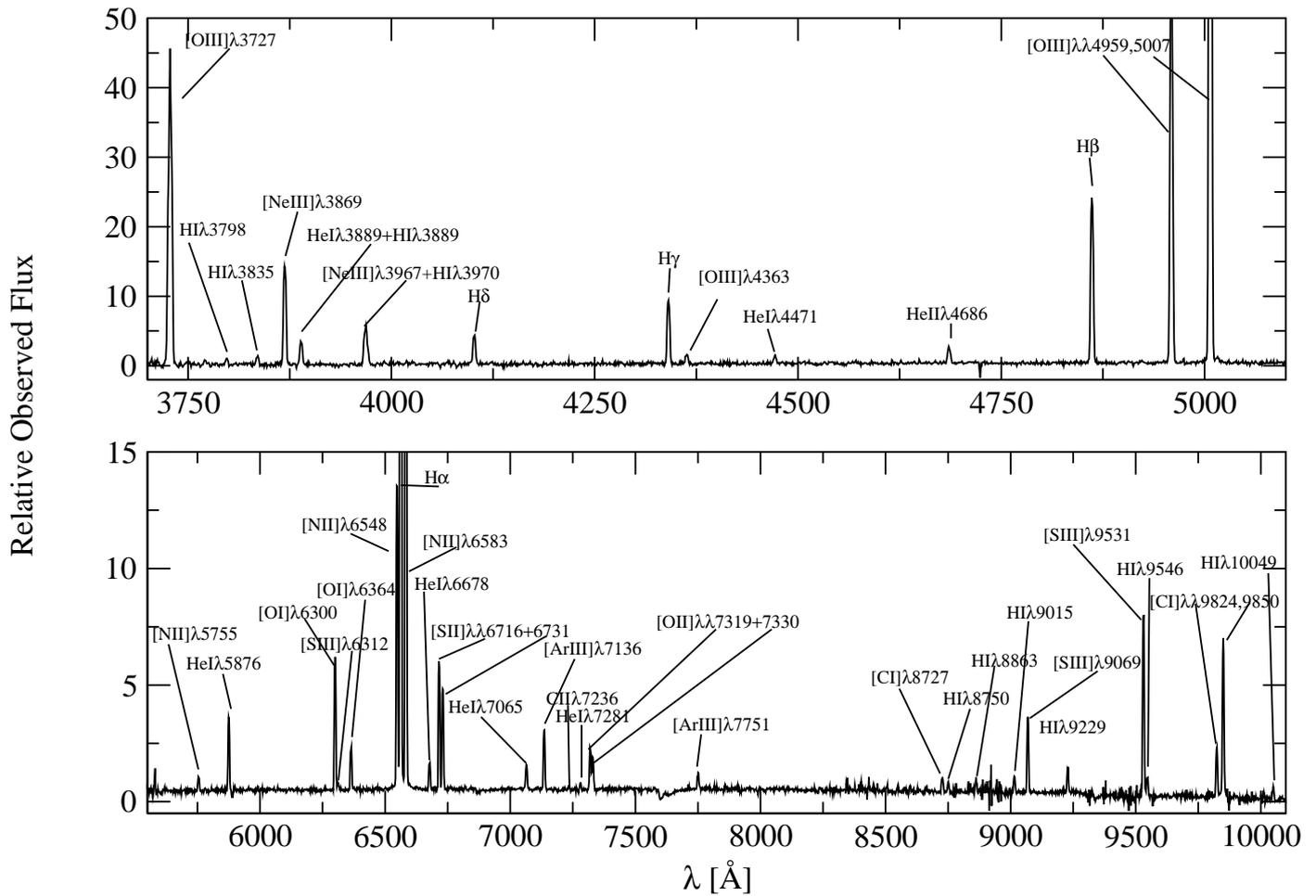}
      \caption{The WHT+ISIS spectrum of IACPN.}
\label{fig:3}
\end{figure*}

A one-dimensional spectrum of the object (both from the WHT and the
NOT data) was extracted summing the nebular spectra along the spatial
axis. The WHT spectrum is shown in Fig.~\ref{fig:3}.

Line fluxes were measured fitting Gaussian profiles using the task
{\sc splot} of IRAF. The errors in the line fluxes were calculated
considering both the statistical errors and errors due to the flux
calibration. The latter were obtained for each line from the error in
the calibration data at the wavelength under question and the local
RMS of the fitted calibration curve, and were added quadratically to
the statistical errors. Resulting line fluxes are listed in 
Tables~\ref{tab:wht} (WHT) and \ref{tab:not} (NOT); fluxes in common in both datasets were found to agree within 3-$\sigma$, and in most cases within 1-$\sigma$. 
The WHT line fluxes were used in the physico-chemical analysis that follows.

\subsection{Reddening}\label{redd}

The reddening law of \cite{fitzpatrick04} for $R_V$ = 3.1, and
  the Balmer line fluxes from Tables~\ref{tab:wht} and \ref{tab:not}
  were used to derive the extinction coefficient $c_\beta$ from both
  the WHT and NOT spectra. For the WHT data, we derive
  $c_{H\beta}=0.7\pm0.1$ and $0.9\pm0.1$ from the H$\gamma$/H$\beta$
  and H$\delta$/H$\beta$ line ratios, respectively. For the NOT, very
  similar values are obtained, namely $c_{H\beta}=0.8\pm0.1$,
  $0.7\pm0.2$, and $0.6\pm0.2$ from the H$\alpha$/H$\beta$,
  H$\gamma$/H$\beta$, and H$\delta$/H$\beta$ ratios, respectively.
 The weighted mean of all measurements is $c_{H\beta}=0.8\pm0.1$,
  i.e. $A_V=1.7\pm0.2$. 

We have adopted this value to deredden the NOT spectrum and,
  separately, the blue and red WHT spectra of the nebula.  Given
  the lack of a precise match of the blue and red parts in the WHT data
  (Sect. 2.4) a final correction to these data was
  applied after dereddening by rescaling the whole red spectrum 
to match the theoretical value
  $I(\mbox{H}\alpha)/I(\mbox{H}\beta)=2.86$ \citep{osterbrock74}.

The extinction derived in this paper is within the errors, albeit
slightly larger, than the value from \cite{mampaso05}, $A_V=1.5\pm0.2$
mag (derived from a noiser H$\alpha$/H$\beta$ ratio and using a
different extinction law).

\subsection{Density and temperature}

The IRAF package {\sc temden} was used to derive the electronic
density and temperature.  The density was derived by iterating the
density calculated from the [S\,{\sc ii}]6716 \AA/6731 \AA~line ratio
and the temperature derived from the [N\,{\sc
    ii}](6548 \AA+6583 \AA)/5755 \AA~line ratio. This leads to
$N_e=$190$\pm$120 cm$^{-3}$, when the errors in the dereddened sulphur line
fluxes are considered. The density error is large as the
[S\,{\sc ii}]6716 \AA/6731 \AA~line ratio is at the upper limit of its
usability for the $N_e$ determination.

Adopting the [S\,{\sc ii}] density, the electronic temperature was
derived from [N\,{\sc ii}](6548 \AA+6583 \AA)/5755 \AA, [O\,{\sc ii}](3726 \AA+3729 \AA)/(7320 \AA+7330 \AA), and [O\,{\sc iii}](4959 \AA+5007 \AA)/4363 \AA~line ratios giving, respectively, 11700$\pm$800 K, 12000$\pm$800 K, and 11700$\pm$300~K.

\subsection{Chemical abundances}

The ionic abundances of He$^+$ and He$^{++}$ and corresponding errors 
were calculated using the equations of \cite{benjamin99}  which include the effects of collisional excitation on He\,{\sc i}
lines. The ionic abundances for the rest of the elements were
calculated using the {\sc nebular} analysis package in IRAF/STSDAS
\citep{shaw94}.
For the ions with the lowest ionisation potentials, O$^+$ ([O\,{\sc ii}] 3727 \AA, 7319 \AA, 7330 \AA), N$^+$ ([N\,{\sc ii}] 5755 \AA, 6548 \AA, 6583 \AA), and
S$^+$ ([S\,{\sc ii}] 6716 \AA + 6731 \AA) the $T_e$[N\,{\sc ii}] was adopted, while for the higher
ionisation potential ions, He$^+$ (He\,{\sc i} 4471 \AA, 5876 \AA, 6678 \AA), He$^{++}$ (He\,{\sc ii} 4686 \AA), O$^{++}$ ([O\,{\sc iii}] 4363 \AA, 4959 \AA, 5007 \AA), Ne$^{++}$ ([Ne\,{\sc iii}] 3869 \AA),
S$^{++}$ ([S\,{\sc iii}] 6312 \AA, 9069 \AA), and Ar$^{++}$ ([Ar\,{\sc iii}] 7135 \AA, 7751 \AA), the $T_e$[O\,{\sc iii}] was used. The
[S\,{\sc ii}] density was adopted for all the ions. The errors were
calculated using the Monte Carlo method, assigning for each variable
(flux, temperature, and density) 100 random values within a normal distribution defined by the corresponding quantity and its 1-$\sigma$ uncertainty and calculating the resulting RMS in the ionic
abundance. The ionic abundances derived from different lines of the same ion agree within 1.5$\sigma$ in all cases. The total elemental abundances were then determined by applying ionisation correction factors (ICFs) following the prescriptions by \cite{kingsburgh94}, and the errors were propagated to include the ionic abundance errors derived above and the errors in the ICF's. The ionic and total abundances with their corresponding errors are given in Table~\ref{tab:abund}.

\begin{table}
\caption{\label{tab:abund} Ionic and total abundances for IACPN. The percentage errors are given within brackets.}  
\centering
 \begin{tabular}{@{}lllll@{}}
 \hline \hline
Ion/Element & Abundance & 12+log(X/H) \\
\hline
He\,{\sc i} & 0.117 (10) & \\
He\,{\sc ii} & 0.009 (7) & \\
{\bf He/H} & {\bf 0.125 (9)} & {\bf 11.10$\pm$0.04} \\
{}[O\,{\sc ii}] & 7.69E-5 (4) & \\
{}[O\,{\sc iii}] & 1.42E-4 (2) & \\
{\it icf}(O) & 1.05 (8) &\\
{\bf O/H} & {\bf 2.30E-4 (7)} & {\bf 8.36$\pm$0.03} \\
{}[N\,{\sc ii}] & 2.07E-05 (0.2) & \\
{\it icf}(N) & 2.99 (8) &\\
{\bf N/H} & {\bf 6.20E-05 (8)} & {\bf 7.79$\pm$0.03} \\
{}[Ar\,{\sc iii}] & 5.72E-07 (1) & \\
{\it icf}(Ar) & 1.87 (22) & \\
{\bf Ar/H} & {\bf 1.07E-06 (22)} & {\bf 6.03$\pm$0.10} \\
{}[S\,{\sc ii}] & 6.14e-07 (18) & \\
{}[S\,{\sc iii}] & 1.50E-06 (3) & \\
{\it icf}(S) & 1.12 (1) &  \\
{\bf S/H} & {\bf 2.37E-06 (6)} & {\bf 6.37$\pm$0.02} \\
{}[Ne\,{\sc iii}] & 5.26E-05 (12) & \\
{\it icf}(Ne) & 1.62 (7) &\\
{\bf Ne/H} & {\bf 8.50E-05 (14)} & {\bf 7.93$\pm$0.06} \\
\hline
\end{tabular}
\end{table}

\section{Distance}\label{dist}

The distance to IACPN was estimated using two different methods. 

\begin{figure}
   \centering
   \includegraphics[width=\linewidth,angle=0]{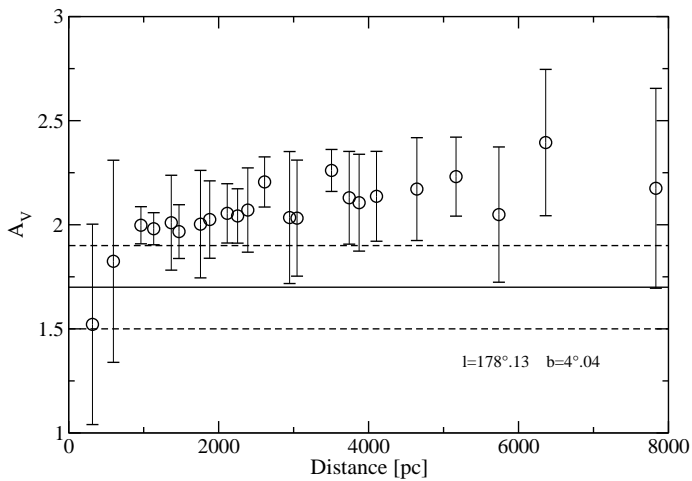}
      \caption{IPHAS extinction -- distance curve (circles) for field stars towards the sightline of IACPN. The horizontal line shows the measured extinction $A_V$ of IACPN flanked by two dashed lines representing $\pm 1\sigma$ error limits.}

\label{fig:4}
\end{figure}

Firstly, we used the extinction--distance technique ($A_v - D$; c.f. \citealt{gathier86}) as implemented by \cite{sale09} for IPHAS data. The $A_v - D$ method is currently the only one which is capable of determining accurate ($\sim 30\%$) distances for a large number of PNe, being independent of statistical assumptions about the physics of the nebulae and/or of their central stars \citep{giammanco11}. GAIA will hopefully improve this situation in the near future by providing hundreds of trigonometric PN distances. The $A_v - D$ curve towards the sightline of IACPN ($l,b = 178.13, -4.04$) was calculated using photometry of 3629 stars in a $10 \arcmin \times 10\arcmin$ box centred on the position of the PN and is shown in Fig.~\ref{fig:4}. Most of the interstellar extinction builds up within the first kiloparsec, beyond which it remains roughly constant at $A_V \sim 2$. We note that this behaviour is consistent with the extinction map of \citet{schlegel98} which gives an integrated Galactic extinction of $A_V = 1.9$ towards this sightline, i.e. close to the asymptotic value shown by Fig.~\ref{fig:4}. The likely reason for such an abrupt asymptoting is that, at the distance of the Perseus arm (the only major dusty structure encountered beyond 1~kpc along this line of sight), a Galactic latitude of $b=-4.04^{\circ}$ corresponds to a distance of $\sim 140$~pc below the midplane, and little cumulative dust obscuration is expected farther out.

According to Fig.~\ref{fig:4}, IACPN could be located at any distance from $<$1 kpc (the formal best fit to the uprising section of the curve) to $\gtrsim$8 kpc (the limit where there are adequate stellar data). The error bars of the stellar and the PN data overlap at $2\sigma$ and the plateau of interstellar extinction along this line of sight precludes a thrustworthy estimation of the extinction distance to this PN. 

Secondly, the relationship between H$\alpha$-surface brightness and nebular radius ($S - r$) of \citet{frew06,frew10} was used. IACPN is almost circular, and its dimensions, measured directly from the H$\alpha$ IPHAS image at the 10 percent of peak-brightness contour, following the procedure of \citet{tylenda03}, are 10.5 $\times$ 9.7 arcsec.

The total H$\alpha$ flux was calculated from the wide-slit H$\beta$ flux (Section 2.2) and the Balmer decrement of 5.5 $\pm$ 0.2, resulting in log~F(H$\alpha$) = $-12.73$ $\pm$ 0.05. The H$\alpha$ flux was also independently estimated from the IPHAS calibrated H$\alpha$ image after subtracting the contribution of the two adjacent [N\,{\sc ii}] lines as measured form the spectrum, yielding log~F(H$\alpha$) = $-12.96$ $\pm$ 0.22, in fair agreement with the spectroscopically determined value. We will adopt the better determined spectroscopic value in the following.

The resulting H$\alpha$ surface brightness is log~S(H$\alpha$) = $-4.00$ $\pm$ 0.06 erg\,cm$^{-2}$\,s$^{-1}$\,sr$^{-1}$. After dereddening this value using $A_{V}$ = 1.7 $\pm$ 0.2 mag and the extinction law of Fitzpatrick (2004) with $R_{V}$ = 3.1, we find the physical radius of the PN to be 0.31\,pc, using the mean H$\alpha$ $S - r$ relation of \cite{frew06}.  The mean-trend equation \citep[][ Frew et al. 2011, in preparation]{frew08} was used, which is applicable to round, optically-thick PN, viz.:

\begin{equation}
\log S(\mbox{H}\alpha) = -3.65(\pm0.08)\log R - 5.34(\pm0.07).
\end{equation}

The resulting distance is 12.8 $\pm$ 3.7 kpc.  The quoted uncertainty in the distance includes the formal dispersion of the relation ($\pm$ 28\%; \citealt{frew08}) which is the dominant error term, added in quadrature with the estimated uncertainties of the integrated H$\alpha$ flux, mean diameter, and reddening of the PN. The corresponding galacto-centric distance is D$_{GC} = 20.8\pm3.8$ kpc if a distance to the Sun of $8.0\pm0.6$ kpc \citep{ghez08} is assumed. At 12.8 kpc from the Sun, IACPN is located about 1 kpc off the plane of the Galactic disk. That is a large height indeed, but we note that the well-known flaring of the outer disk \citep[see e.g.][]{gyuk99} allows a high probability of thin disk objects being located even at this offset from the mid-plane. Using the \citet{gyuk99} prescription, the thin disk scale height at D$_{GC}$ = 20.8 kpc is in excess of 1 kpc (even if the solar neighbourhood scale height is reduced in line with the \citet{juric08} result).

We will adopt for IACPN the $S - r$ distance of 12.8 kpc.
At D$_{GC}$=20.8 kpc, it would be one of the most distant planetary nebula from the Galactic Centre for which a distance has been determined (c.f. \citealt{acker94,ssv08}) and the farthest PN from the Galactic Centre with measured abundances (\citealt{maciel09,henry10}).

\section{Discussion}\label{disc}

IACPN is a low density planetary nebula with an emission line spectrum typical of a moderate excitation Type II PN \citep[Class 3 of][]{dopita90}. This, together with its strong stratification, apparent from He\,{\sc ii} to O\,{\sc i} (Fig.~\ref{fig:2}), suggests that it is composed of a moderately massive nebula hosting a relatively hot star. The ionised hydrogen mass of the nebula, derived from its H$\beta$ flux, electron temperature, radius, and distance, is M= 0.4 M$_{\odot}$. The nebular expansion velocity (17 km s$^{-1}$) and the physical radius (0.31 pc at the distance of 12.8 kpc) lead to a kinematic age of 17800 yr. Such a large kinematical age would point to a low mass central star if a luminosity larger than $\sim300-500$ L$_{\odot}$ is required \citep[cf.][]{blocker95}. To crudely estimate the central star mass, we follow the prescriptions of \citet{maciel10} for a N/O abundance ratio as given in Table~\ref{tab:abund}, finding m$_{cs} = 0.6$ M$_{\odot}$ and an age for the progenitor star of $\sim$4 Gyr (if Case A from \citealt{maciel10} is assumed). However, given the several uncertain parameters involved (central star luminosity and temperature, kinematical vs. real age), no precise estimate of the stellar mass can be derived. 

We have ran several simple Cloudy \citep{ferland98} photoionisation models using
spherical or shell geometry and the model atmospheres of \citet{rauch03}. We assumed a nebula with an outer radius of 0.31 pc, hydrogen density of 200 cm$^{-3}$, and chemical abundances 0.3 dex below those labelled in Cloudy as ``planetary nebula''. This implies that the input models have 12+log(O/H)=8.34 and He/H=0.1, close to our measured values. On the other hand, the model abundances of N, Ar, and S are ~0.1-0.3 dex above our derived values, whereas the model abundance of Ne is 0.2 dex below the value listed in Table~\ref{tab:abund}. Since the abundances of these elements are more uncertain, and since our comparison is based on the He and O ions, this should be a good approximation for our purposes. We considered the observed intensities of the He\,{\sc i}, He\,{\sc ii}, [O\,{\sc i}], [O\,{\sc ii}], and [O\,{\sc iii}] lines,
as given in Table~\ref{tab:wht}, and found that high stellar temperatures (T$_{eff} \gtrsim 10^5$ K) are needed to account for the observed intensity of the He\,{\sc ii} 4686 \AA~line, in agreement with the results presented above. Models with those effective temperatures also reproduce the intensities of all the aforementioned lines. The radiation-bounded case requires a luminosity for the ionising source L$_{cs} \sim 100$ L$_{\odot}$, being therefore consistent with a low mass star on the cooling track. As a matter of fact, similar models for radiation-bounded nebulae as small as 0.01 pc (corresponding to extinction distances smaller than 1 kpc; see Section 4) would imply central star luminosities around 0.001 L$_{\odot}$, i.e., much lower than the expected luminosity range of 10-1000 L$_{\odot}$ for stars on the cooling track. This renders implausible models using very short distances for this PN. 

We also investigated the [O\,{\sc iii}] luminosity of IACPN to compare it to the bright-end cutoff of the PN luminosity function \citep[PNLF:][]{Ciardullo10}. We used the long-slit integrated H$\beta$ flux and the [O\,{\sc iii}]/H$\beta$ ratio of 7.3 to derive an integrated [O\,{\sc iii}] flux, $F$(5007) = 2.43 $\times$ 10$^{-13}$ erg\,cm$^{-2}$\,s$^{-1}$.  Correcting for extinction, and following \citet{jacoby89}, we derive a PN absolute magnitude, $M_{5007}$ = +0.3$^{+0.7}_{-0.6}$ at the adopted distance.  The PN is thus $\sim$4.8 magnitudes down the PNLF from the bright cut-off magnitude, $M^{\ast}$ = $-4.48$. This is consistent with the evolved nature of the PN, and the position of the (unobserved) central star on the WD cooling track.

In summary, all physical data at hand on IACPN are consistent with a distant, aged, radiation-bounded Type II planetary nebula excited by a low mass central star with relatively high temperature and having already reached its cooling track.

With respect to the chemistry, the O/H abundance of 8.36$\pm$0.03 (Table~\ref{tab:abund}) is interestingly high for the large galactocentric distance, 20.8 kpc, and makes IACPN a key target for studies of the abundance gradient in the Milky Way. It is generally agreed that there is {\it some} negative PNe metallicity gradient but the consensus seems to finish there. Values for the best measured element, oxygen, range from $\Delta$log(O/H)/D$_{GC} = -0.01$ \citep{pottasch06} to $-0.08$ \citep{stanghellini06} with most recent values still disagreeing by a factor of two, well above the uncertainties quoted by the authors: c.f. $-0.023\pm 0.006$ and $-0.058\pm 0.006$ from \citet{stanghellini10} and \citet{henry10}, respectively. Similar confusion hampers the possible existence of a variable slope, i.e. if the gradient steepens \citep{stanghellini10} or flattens \citep{maciel10b} outwards along the Galactic equatorial plane, and also on its secular temporal behaviour, either null \citep{henry10} or steeping \citep{stanghellini10} or flattening with time \citep{maciel10b}. Perhaps the most sensible conclusion comes from \citet{henry10}: ``Essentially we have reached a confusion limit regarding the abundance gradient as derived using PNe.". One of the major sources of this confusion is the unsettled PNe distance determination problem allied with the extreme paucity of nebulae at large galactocentric distances. In fact, in the most recent works, those by \cite{henry10} and \cite{stanghellini10}, only two PNe with $D_{GC}> 17 kpc$ are included in each sample (K3-64, K3-70, and M1-9, K3-66, respectively), all four PNe having mediocrily determined O abundances (quoted errors from 0.2 to 0.5 dex) and distances. However, these distant nebulae are obviously critical for the gradient determination.

In this context, IACPN provides a valuable data point located at the outer reaches of the Galaxy, and its well-constrained O abundance do support models with a flatter gradient toward the external disk. An intriguing possibility is that IACPN and perhaps  also the other four PNe with D$_{GC}$ $>$17~kpc
might even have an extragalactic origin, if indeed the Monoceros Ring proposed
to be located at these radii \citep[eg.][]{conn08} is attributable to a satellite accretion event.  A consensus on this has not yet been reached \citep[eg.][]{hammersley10}. But either way it is clear that the properties of the outer Galactic disc are still in need of much better observational definition.

\section{Conclusions}\label{conc}

We are exploring the new IPHAS planetary nebula candidates \citep{viironen09b} towards the Galactic Anticentre in order to better sample the Galactic abundance gradient at large galactocentric distances.
In this paper, the discovery of a new PN located at only 4 degrees from the Galactic Anticentre, IPHASX J052531.19+281945.1, is presented, its physical and
chemical properties are studied and its distance determined. The object turned out
to be a low-density, distant, and aged PN. It is the Galactic PN with the largest galactocentric distance (20.8$\pm$3.8 kpc) where the chemical abundances have been measured, and this, combined with its relatively high oxygen abundance, 12 + log(O/H) = 8.36$\pm$0.03, points to a flattening of the Galactic
abundance gradient towards the outer Galaxy rather than to a
linearly decreasing oxygen abundance  \citep[e.g.][and references therein]{maciel09}.

To properly study the behaviour of the abundance gradient at the outer Galaxy, reasonably accurate abundance measurements on a sample of objects at large galactocentric distances are needed. We have recently discovered in the IPHAS images about 180 new faint PN candidates located in a $30^{\circ} \times 10^{\circ}$ region around the Anticentre \citep{viironen09b}. Further abundance
measurements of confirmed candidates in this region would help to build a robust chemical gradient determination and
reach firmer conclusions concerning its flattening.

\begin{acknowledgements}
This paper makes use of data obtained as part of IPHAS carried out at the
INT, as well as WHT, NOT/ALFOSC and SPM2.1m spectroscopic data. The INT and
WHT are operated by the Isaac Newton Group, whereas NOT is operated jointly by Denmark, Finland, Iceland, Norway, and Sweden, both are at the
island of La Palma in the Spanish Observatorio del Roque de los
Muchachos of the Instituto de Astrof\'{i}sica de Canarias. ALFOSC is owned by
the Instituto de Astrof\'{i}sica de Andalucia (IAA) and operated at the
NOT under agreement between IAA and the NBIfAFG
of the Astronomical Observatory of Copenhagen. SPM 2.1m is
operated by UNAM at the OAN of SPM, Mexico. All IPHAS data are
processed by the Cambridge Astronomical Survey Unit, at the Institute
of Astronomy in Cambridge. L. Sabin is grateful for receiving a UNAM postdoctoral fellowship and is partially supported by PAPIIT-UNAM grant IN109509 (Mexico). Support for S. E. Sale is provided by MIDEPLAN's Programa Inicativa Cient\'{i}fica Milenio through grant P07-021-F, awarded to The Milky Way Millennium Nucleus. We thank J.~A. L\'opez and H. Riesgo for kindly acquiring and reducing the SPM MEZCAL spectra for us, NOT staff (Pilar Monta\~n\'es Rodr\'iguez, Tapio Pursimo) for obtaining the NOT spectrum, and Q. A. Parker for discussions. Finally, we thank the anonymous referee for helping us to substantially improve the article.
\end{acknowledgements}

\bibliographystyle{aa}
\bibliography{14897bib.bib}

\begin{thebibliography}{40}
\expandafter\ifx\csname natexlab\endcsname\relax\def\natexlab#1{#1}\fi

\bibitem[{{Acker} {et~al.}(1994){Acker}, {Ochsenbein}, {Stenholm}, {Tylenda},
  {Marcout}, \& {Schohn}}]{acker94}
{Acker}, A., {Ochsenbein}, F., {Stenholm}, B., {et~al.} 1994, VizieR Online
  Data Catalog, 5084

\bibitem[{{Benjamin} {et~al.}(1999){Benjamin}, {Skillman}, \&
  {Smits}}]{benjamin99}
{Benjamin}, R.~A., {Skillman}, E.~D., \& {Smits}, D.~P. 1999, \apj, 514, 307

\bibitem[{{Bl\"ocker}(1995)}]{blocker95}
{Bl\"ocker}, T. 1995, \aap, 299, 755

\bibitem[{{Ciardullo}(2010)}]{Ciardullo10}
{Ciardullo}, R. 2010, \pasa, 27, 149

\bibitem[{{Conn} {et~al.}(2008){Conn}, {Lane}, {Lewis}, {Irwin}, {Ibata},
  {Martin}, {Bellazzini}, \& {Tuntsov}}]{conn08}
{Conn}, B.~C., {Lane}, R.~R., {Lewis}, G.~F., {et~al.} 2008, \mnras, 390, 1388

\bibitem[{{Dopita} \& {Meatheringham}(1990)}]{dopita90}
{Dopita}, M.~A. \& {Meatheringham}, S.~J. 1990, \apj, 357, 140

\bibitem[{{Drew} {et~al.}(2005){Drew}, {Greimel}, {Irwin}, {Aungwerojwit},
  {Barlow}, {Corradi}, {Drake}, {G{\"a}nsicke}, {Groot}, {Hales}, {Hopewell},
  {Irwin}, {Knigge}, {Leisy}, {Lennon}, {Mampaso}, {Masheder}, {Matsuura},
  {Morales-Rueda}, {Morris}, {Parker}, {Phillipps}, {Rodriguez-Gil}, {Roelofs},
  {Skillen}, {Sokoloski}, {Steeghs}, {Unruh}, {Viironen}, {Vink}, {Walton},
  {Witham}, {Wright}, {Zijlstra}, \& {Zurita}}]{drew05}
{Drew}, J.~E., {Greimel}, R., {Irwin}, M.~J., {et~al.} 2005, \mnras, 362, 753

\bibitem[{{Ferland} {et~al.}(1998){Ferland}, {Korista}, {Verner}, {Ferguson},
  {Kingdon}, \& {Verner}}]{ferland98}
{Ferland}, G.~J., {Korista}, K.~T., {Verner}, D.~A., {et~al.} 1998, \pasp, 110,
  761

\bibitem[{{Fitzpatrick}(2004)}]{fitzpatrick04}
{Fitzpatrick}, E.~L. 2004, in Astronomical Society of the Pacific Conference
  Series, Vol. 309, Astrophysics of Dust, ed. {A.~N.~Witt, G.~C.~Clayton, \&
  B.~T.~Draine}, 33

\bibitem[{{Frew}(2008)}]{frew08}
{Frew}, D.~J. 2008, {PhD thesis} (Macquarie University)

\bibitem[{{Frew} \& {Parker}(2006)}]{frew06}
{Frew}, D.~J. \& {Parker}, Q.~A. 2006, in IAU Symposium, Vol. 234, Planetary
  Nebulae in our Galaxy and Beyond, ed. M.~J. {Barlow} \& R.~H. {M{\'e}ndez},
  49--54

\bibitem[{{Frew} \& {Parker}(2010)}]{frew10}
{Frew}, D.~J. \& {Parker}, Q.~A. 2010, \pasa, 27, 129

\bibitem[{{Gathier} {et~al.}(1986){Gathier}, {Pottasch}, \& {Pel}}]{gathier86}
{Gathier}, R., {Pottasch}, S.~R., \& {Pel}, J.~W. 1986, \aap, 157, 171

\bibitem[{{Ghez} {et~al.}(2008){Ghez}, {Salim}, {Weinberg}, {Lu}, {Do}, {Dunn},
  {Matthews}, {Morris}, {Yelda}, {Becklin}, {Kremenek}, {Milosavljevic}, \&
  {Naiman}}]{ghez08}
{Ghez}, A.~M., {Salim}, S., {Weinberg}, N.~N., {et~al.} 2008, \apj, 689, 1044

\bibitem[{{Giammanco} {et~al.}(2011){Giammanco}, {Sale}, {Corradi}, {Barlow},
  {Viironen}, {Sabin}, {Santander-Garc{\'{\i}}a}, {Frew}, {Greimel},
  {Miszalski}, {Phillipps}, {Zijlstra}, {Mampaso}, {Drew}, {Parker}, \&
  {Napiwotzki}}]{giammanco11}
{Giammanco}, C., {Sale}, S.~E., {Corradi}, R.~L.~M., {et~al.} 2011, \aap, 525,
  A58+

\bibitem[{{Gonz{\'a}lez-Solares} {et~al.}(2008){Gonz{\'a}lez-Solares},
  {Walton}, {Greimel}, {Drew}, {Irwin}, {Sale}, {Andrews}, {Aungwerojwit},
  {Barlow}, {van den Besselaar}, {Corradi}, {G{\"a}nsicke}, {Groot}, {Hales},
  {Hopewell}, {Hu}, {Irwin}, {Knigge}, {Lagadec}, {Leisy}, {Lewis}, {Mampaso},
  {Matsuura}, {Moont}, {Morales-Rueda}, {Morris}, {Naylor}, {Parker}, {Prema},
  {Pyrzas}, {Rixon}, {Rodr{\'{\i}}guez-Gil}, {Roelofs}, {Sabin}, {Skillen},
  {Suso}, {Tata}, {Viironen}, {Vink}, {Witham}, {Wright}, {Zijlstra}, {Zurita},
  {Drake}, {Fabregat}, {Lennon}, {Lucas}, {Mart{\'{\i}}n}, {Phillipps},
  {Steeghs}, \& {Unruh}}]{gonzalez-solares08}
{Gonz{\'a}lez-Solares}, E.~A., {Walton}, N.~A., {Greimel}, R., {et~al.} 2008,
  \mnras, 707

\bibitem[{{Gyuk} {et~al.}(1999){Gyuk}, {Flynn}, \& {Evans}}]{gyuk99}
{Gyuk}, G., {Flynn}, C., \& {Evans}, N.~W. 1999, \apj, 521, 190

\bibitem[{{Hammersley} \& {Lopez-Corredoira}(2010)}]{hammersley10}
{Hammersley}, P.~L. \& {Lopez-Corredoira}, M. 2010, ArXiv e-prints

\bibitem[{{Henry} {et~al.}(2010){Henry}, {Kwitter}, {Jaskot}, {Balick},
  {Morrison}, \& {Milingo}}]{henry10}
{Henry}, R.~B.~C., {Kwitter}, K.~B., {Jaskot}, A.~E., {et~al.} 2010, \apj, 724,
  748

\bibitem[{{Jacoby}(1989)}]{jacoby89}
{Jacoby}, G.~H. 1989, \apj, 339, 39

\bibitem[{{Juri{\'c}} {et~al.}(2008){Juri{\'c}}, {Ivezi{\'c}}, {Brooks},
  {Lupton}, {Schlegel}, {Finkbeiner}, {Padmanabhan}, {Bond}, {Sesar},
  {Rockosi}, {Knapp}, {Gunn}, {Sumi}, {Schneider}, {Barentine}, {Brewington},
  {Brinkmann}, {Fukugita}, {Harvanek}, {Kleinman}, {Krzesinski}, {Long},
  {Neilsen}, {Nitta}, {Snedden}, \& {York}}]{juric08}
{Juri{\'c}}, M., {Ivezi{\'c}}, {\v Z}., {Brooks}, A., {et~al.} 2008, \apj, 673,
  864

\bibitem[{{Kingsburgh} \& {Barlow}(1994)}]{kingsburgh94}
{Kingsburgh}, R.~L. \& {Barlow}, M.~J. 1994, \mnras, 271, 257

\bibitem[{{Kwok}(2000)}]{kwok00}
{Kwok}, S. 2000, {The Origin and Evolution of Planetary Nebulae} (Cambridge ;
  New York : Cambridge University Press, 2000.~(Cambridge astrophysics series ;
  33))

\bibitem[{{Maciel} \& {Costa}(2009)}]{maciel09}
{Maciel}, W.~J. \& {Costa}, R.~D.~D. 2009, in IAU Symposium, ed. {J.~Andersen,
  J.~Bland-Hawthorn, \& B.~Nordstr{\"o}m}, Vol. 254, 38

\bibitem[{{Maciel} \& {Costa}(2010)}]{maciel10b}
{Maciel}, W.~J. \& {Costa}, R.~D.~D. 2010, in IAU Symposium, ed. {K.~Cunha,
  M.~Spite, \& B.~Barbuy}, Vol. 265, 317--324

\bibitem[{{Maciel} {et~al.}(2010){Maciel}, {Costa}, \& {Idiart}}]{maciel10}
{Maciel}, W.~J., {Costa}, R.~D.~D., \& {Idiart}, T.~E.~P. 2010, \aap, 512, A19+

\bibitem[{{Mampaso} {et~al.}(2005){Mampaso}, {Viironen}, {Corradi},
  {Rodr{\'{\i}}guez}, {Drew}, {Greimel}, \& {Irwin}}]{mampaso05}
{Mampaso}, A., {Viironen}, K., {Corradi}, R.~L.~M., {et~al.} 2005, in American
  Institute of Physics Conference Series, Vol. 804, Planetary Nebulae as
  Astronomical Tools, ed. {R.~Szczerba, G.~Stasi{\'n}ska, \& S.~K.~Gorny},
  14--14

\bibitem[{{Oke}(1990)}]{oke90}
{Oke}, J.~B. 1990, \aj, 99, 1621

\bibitem[{{Osterbrock}(1974)}]{osterbrock74}
{Osterbrock}, D.~E. 1974, {Astrophysics of gaseous nebulae} (W.~H.~Freeman and
  Co., 1974.~263 p.)

\bibitem[{{Pottasch} \& {Bernard-Salas}(2006)}]{pottasch06}
{Pottasch}, S.~R. \& {Bernard-Salas}, J. 2006, \aap, 457, 189

\bibitem[{{Rauch}(2003)}]{rauch03}
{Rauch}, T. 2003, \aap, 403, 709

\bibitem[{{Sale} {et~al.}(2009){Sale}, {Drew}, {Unruh}, {Irwin}, {Knigge},
  {Phillipps}, {Zijlstra}, {G{\"a}nsicke}, {Greimel}, {Groot}, {Mampaso},
  {Morris}, {Napiwotzki}, {Steeghs}, \& {Walton}}]{sale09}
{Sale}, S.~E., {Drew}, J.~E., {Unruh}, Y.~C., {et~al.} 2009, \mnras, 392, 497

\bibitem[{{Schlegel} {et~al.}(1998){Schlegel}, {Finkbeiner}, \&
  {Davis}}]{schlegel98}
{Schlegel}, D.~J., {Finkbeiner}, D.~P., \& {Davis}, M. 1998, \apj, 500, 525

\bibitem[{{Shaw} \& {Dufour}(1994)}]{shaw94}
{Shaw}, R.~A. \& {Dufour}, R.~J. 1994, in Astronomical Society of the Pacific
  Conference Series, Vol.~61, Astronomical Data Analysis Software and Systems
  III, ed. {D.~R.~Crabtree, R.~J.~Hanisch, \& J.~Barnes}, 327

\bibitem[{{Stanghellini} {et~al.}(2006){Stanghellini}, {Guerrero}, {Cunha},
  {Manchado}, \& {Villaver}}]{stanghellini06}
{Stanghellini}, L., {Guerrero}, M.~A., {Cunha}, K., {Manchado}, A., \&
  {Villaver}, E. 2006, \apj, 651, 898

\bibitem[{{Stanghellini} \& {Haywood}(2010)}]{stanghellini10}
{Stanghellini}, L. \& {Haywood}, M. 2010, \apj, 714, 1096

\bibitem[{{Stanghellini} {et~al.}(2008){Stanghellini}, {Shaw}, \&
  {Villaver}}]{ssv08}
{Stanghellini}, L., {Shaw}, R.~A., \& {Villaver}, E. 2008, \apj, 689, 194

\bibitem[{{Tylenda} {et~al.}(2003){Tylenda}, {Si{\'o}dmiak}, {G{\'o}rny},
  {Corradi}, \& {Schwarz}}]{tylenda03}
{Tylenda}, R., {Si{\'o}dmiak}, N., {G{\'o}rny}, S.~K., {Corradi}, R.~L.~M., \&
  {Schwarz}, H.~E. 2003, \aap, 405, 627

\bibitem[{{Viironen} {et~al.}(2009{\natexlab{a}}){Viironen}, {Greimel},
  {Corradi}, {Mampaso}, {Rodr{\'{\i}}guez}, {Sabin}, {Delgado-Inglada}, {Drew},
  {Giammanco}, {Gonz{\'a}lez-Solares}, {Irwin}, {Miszalski}, {Parker},
  {Rodr{\'{\i}}guez-Flores}, \& {Zijlstra}}]{viironen09b}
{Viironen}, K., {Greimel}, R., {Corradi}, R.~L.~M., {et~al.}
  2009{\natexlab{a}}, \aap, 504, 291

\bibitem[{{Viironen} {et~al.}(2009{\natexlab{b}}){Viironen}, {Mampaso},
  {Corradi}, {Rodr{\'{\i}}guez}, {Greimel}, {Sabin}, {Sale}, {Unruh},
  {Delgado-Inglada}, {Drew}, {Giammanco}, {Groot}, {Parker}, {Sokoloski}, \&
  {Zijlstra}}]{viironen09a}
{Viironen}, K., {Mampaso}, A., {Corradi}, R.~L.~M., {et~al.}
  2009{\natexlab{b}}, \aap, 502, 113

\end{thebibliography}
\end{document}